\documentclass[aps,pra,a4paper,twocolumn,10pt,longbibliography,superscriptaddress]{revtex4-1}
\usepackage{amsmath}
\usepackage{amssymb}
\usepackage{mathtools}
\usepackage{graphics}
\usepackage{comment}
\usepackage{tikz}
\usepackage[normalem]{ulem}

\newcommand{\beq}{\begin{equation}}
\newcommand{\eeq}{\end{equation}}
\newcommand{\bei}{\begin{itemize}}
\newcommand{\eei}{\end{itemize}}
\newcommand{\ben}{\begin{enumerate}}
\newcommand{\een}{\end{enumerate}}

\newcommand{\be}{{\mathbf e}}

\newcommand{\ket}[1]{|#1\rangle}
\newcommand{\bra}[1]{\langle #1|}

\usepackage{hyperref}
\hypersetup{
   pdfpagemode=UseNone, 
   pdfstartpage=1,
   pdfmenubar=true,
   pdftoolbar=true,
   colorlinks = true,
   linkcolor=blue,
   citecolor=blue,
   urlcolor=blue,
   bookmarksopen=false
 }

\definecolor{darkblue}{rgb}{0.,0.24,0.8}
\definecolor{britishracinggreen}{rgb}{0.0, 0.26, 0.15}
\definecolor{darkgreen}{rgb}{0,0.60,.2}

\def\be{\begin{equation}}
\def\ee{\end{equation}}

\begin{document}
\title{Boson-Gutzwiller Quantum liquids on a lattice}
\author{Daniel Pérez-Cruz}
\affiliation{Departamento de F\'isica and IUdEA,
Universidad de La Laguna, La Laguna, 38206, Tenerife (Spain)}
\author{Manuel Valiente}
\affiliation{Departamento de Física, CIOyN, Universidad de Murcia, 30071 Murcia (Spain)}

\date{\today}
	
\begin{abstract}
We consider one-dimensional, interacting spinless bosons on a tight-binding lattice described by the Bose-Hubbard model. Besides attractive on-site two-body interactions, we include a three-body repulsive term such that the competition between these two forces allows the formation of self-bound liquid states. We investigate the properties of this system using the Gutzwiller approximation, showing that, indeed, this mean-field approach also supports liquid states. We find that for densities lower than the equilibrium density, the Gutzwiller method, and other mean-field approaches -- such as the Gross-Pitaevskii theory -- feature a sharp transition to the vacuum state. This, however, is avoided by considering local minima of the functional in the standard manner. We also study the excitation spectrum, and calculate the speed of sound, in full agreement with the usual expression obtained from the thermodynamic equation of state. We study their corresponding quantum droplets variationally and find that the results behave in accordance with the one-dimensional liquid drop model. 
\end{abstract}
\maketitle

\section{Introduction}
A quantum liquid, strictly speaking, is defined as a spatially homogeneous system of interacting particles at temperatures sufficiently low that the effects of quantum statistics are important \cite{pines1966theory}. 
These systems provide ideal playgrounds for exploring many-body physics, quantum phase transitions, and emergent phenomena that arise from strong interactions. The study of quantum liquids, meaning those that form self-bound states -- droplets -- in vacuum, and feature equilibrium properties typical of a liquid in the thermodynamic limit, has gained significant attention due to advances in ultracold atom experiments, which allow for precise control over interactions, confinement, and dimensionality \cite{cabrera2018,schmitt2016,ferrier2016,ferioli2019,bottcher2019,semeghini2018}. 

In particular, we are interested in the possibility of confining these systems to one-dimensional geometries. This enhances interaction effects, making them highly controllable platforms for investigating strongly correlated matter. As a result, new avenues have been opened for studying quantum droplets and novel liquid-like phases of matter \cite{Petrov2015,Petrov2016,Morera2020b,Morera2022a}.

Here we include two- and three-body interactions between bosons to stabilize a zero-temperature liquid on a lattice \cite{Morera2022a}. Three-body processes are always present in many-body systems. Still, their origin may be either a true three-body interparticle potential or an emergent residual interaction from the off-shell structure of the two-body potential \cite{Valiente2019,Hammer2013,Sowinski2014,pricoupenko2019}. In one-dimensional geometries, the competition between mean-field repulsion and beyond-mean-field attraction results in the formation of quantum liquid droplets, marking a fundamental advancement in our understanding of low-dimensional quantum fluids.

Even though mean-field methods perform better in higher dimensions, their application to one-dimensional systems still gives successful results. The bosonic Gutzwiller ansatz provides valuable insights into strongly correlated bosonic systems and bridges more sophisticated numerical techniques and experimental observations \cite{Luhmann2013,Carusotto2003,Buonsante2009}. Its simplicity and adaptability make it a valuable tool in the study of ultracold atoms in optical lattices. 

In this work, we make use of the Gutzwiller ansatz to variationally study both ground state properties and low-energy excitations of a Bose-Hubbard model \cite{Fisher1989,Jaksch1998} with two- and three-body contact interactions, and show that it does support a zero-temperature liquid state. We also study the formation of quantum droplets in finite systems and analyze their properties as a function of the number of particles.

\section{Model and method}
We consider ultracold bosons on a tight-binding lattice interacting via contact two- and three-body interactions \cite{Cazalilla2011}. This system is known to support a quantum liquid (droplet) ground state in the thermodynamic limit (vacuum), for combined attractive two-body and repulsive three-body potentials in one dimension \cite{Morera2022a}. The corresponding Hamiltonian is given by
\begin{equation}
    H=-J\sum_{j}\left[b_j^{\dagger}b_{j+1}+b_{j+1}^{\dagger}b_j-2n_j\right]+H_{\mathrm{I}},\label{eq:hamiltonian}
\end{equation}
where $J$ ($>0$) is the tunnelling rate, $b_j$ ($b_j^{\dagger}$) is the bosonic annihilation (creation) operator at site $j$, $n_j=b_j^{\dagger}b_j$ is the corresponding number operator and we have set the lattice spacing $d=1$. The interaction is given by
\begin{equation}
    H_{\mathrm{I}}=\frac{U}{2}\sum_jn_j(n_j-1)\left[1+\frac{W}{3U}(n_j-2)\right],
\end{equation}
where $U$ ($<0$) and $W$ ($>0$) are, respectively, the two- and three-body interaction strengths.

We shall use the Boson-Gutzwiller mean-field method at zero temperature to study the thermodynamic equation of state and low-energy excitations, as well as the droplets of the system in vacuum. The method is variational, and the underlying wave function --  the Gutzwiller Ansatz -- takes the following form \cite{Bloch2008,Rokhsar1991,Gutzwiller1963, Krauth1992}
\begin{equation}
\label{eq:Gutz_gs}
    \ket{\Psi} = \bigotimes_j \ket{\phi_j}, \hspace{10pt}   \ket{\phi_j} = \sum_{n=0}^{\infty}  a_{n}(j) \ket{n_j},
\end{equation}
where $\ket{n_j}$ is the Fock state with $n$ particles at site $j$, and $a_{n}(j)$ are the expansion coefficients. Since the Gutzwiller wave function, Eq.~(\ref{eq:Gutz_gs}), does not have a fixed particle number, we work in the grand canonical ensemble, that is, we use the grand potential $\Omega$ in place of the Hamiltonian $H$, with $\Omega=H-\mu N$, where $N=\sum_j n_j$ is the total number operator, and $\mu$ is the chemical potential.

The simplicity of the Gutzwiller Ansatz allows us to study a wide range of system properties in a simple and clear manner. This approach is equivalent (at zero temperature) to the following decoupling approximation 
\begin{equation}
    b^\dagger_ib_j \approx \langle b_i^\dagger \rangle b_j + \langle b_j \rangle  b^\dagger_i -\langle b_j \rangle \langle b_i^\dagger \rangle,\label{eq:decoupling}
\end{equation}
as was shown in Ref.~\cite{Sheshadri1993}. We note that the bosonic Gutzwiller approximation can also be extended to account for quantum fluctuations \cite{Caleffi2021,Jolicoeur1991} around the mean-field solution.

The relevant local quantity that characterizes the behaviour of the system is the superfluid order parameter $\psi_j$ at site $j$, defined as:
\begin{equation}
    \label{eq:order_parameter}
    \psi_j = \langle b_j \rangle =  \sum_{n=0}^{\infty} a_{n}^*(j) a_{n-1}(j) \sqrt{n}.
\end{equation}
The superfluid density at site $j$ is given by $|\psi_j|^2$, and is bounded from above by the total number of particles at site $j$, given by
\begin{equation}
    \langle n_j \rangle = \sum_{n=0}^{\infty} n |a_{n}(j)|^2.
\end{equation}

\section{Homogeneous system and equation of state}
We first consider a homogeneous system with uniform density, that is, in the thermodynamic limit. This brings a major simplification of the problem since all local wave functions $\ket{\phi_j}$ in Eq.~(\ref{eq:Gutz_gs}) become identical (site-independent). Therefore, we define, for all $j$,
\begin{equation}
\ket{\phi_j} \equiv \ket{\phi} = \sum_n  a_{n} \ket{n}.\label{eq:phi}
\end{equation}
This reduces the minimization of the grand potential $\Omega$ to a single-site problem, since
\begin{equation}
    \bra{\Psi} \Omega\ket{\Psi} = \sum_j \bra{\phi_j}\Omega \ket{\phi_j}\equiv \sum_{j}\bra{\phi}\Omega_j\ket{\phi},
\end{equation}
where the single-site grand canonical Hamiltonian $\Omega_j$ has the form
\begin{align}
    \Omega_j &= -2J\left[b_j^{\dagger}\psi+\psi^*b_j\right]+(2J-\mu)n_j + 2J|\psi|^2\nonumber \\
    &+\frac{U}{2}n_j(n_j-1)\left[1+\frac{W}{3U}(n_j-2)\right].\label{eq:single_site_ham}
\end{align}

\begin{figure}[b]
    \centering
    \includegraphics[width=\columnwidth]{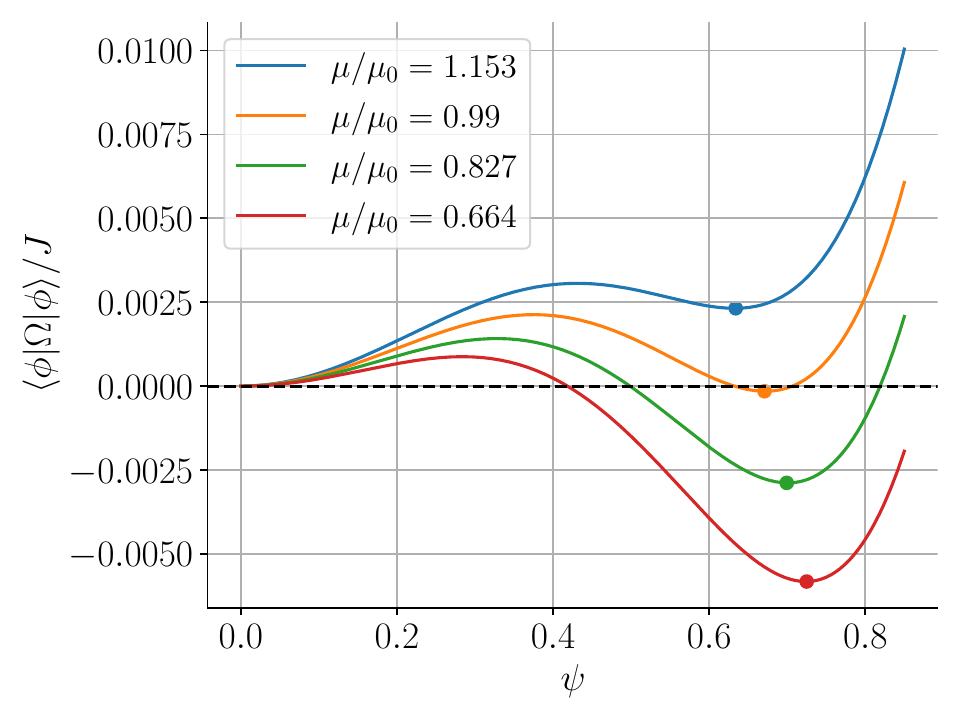}
    \caption{Energy landscape of the grand canonical function in terms of the order parameter $\psi$. The dots represent the local minimum that gives us the non-zero solution for the system's ground state. It is clear that, as the equilibrium density is crossed the vacuum becomes, energetically, the most favorable solution. The parameters of the Hamiltonian (\ref{eq:single_site_ham}) are $U/J = -0.33$, $W/J = 1.39$.}
    \label{fig:order_parameter}
\end{figure}

To see whether a zero-temperature quantum liquid exists, within the Gutzwiller approximation, we minimize the single-site grand canonical Hamiltonian with fixed values of $\mu$. In all cases, we find that the global minimum is attained with non-zero densities $\rho$ when these are higher than the equilibrium density $\rho_0$. If we call the spinodal density $\rho_*$, then for $\rho_*<\rho<\rho_0$, the state that analytically continues the zero-temperature equation of state gives no longer the global minimum, but becomes a local one (see Fig.~\ref{fig:order_parameter}). This is not a feature particular to the Gutzwiller approximation, but common to other mean-field theories. The same issue arises with Gross-Pitaevskii equations (GPE) for quantum liquids: their solutions are built by {\it locally} minimizing the grand canonical functional. As an example, the functional from which the GPE in Ref.~\cite{Astrakharchik2018} by Astrakharchik and Malomed (AM from now on) is obtained, is given, in their units, by
\begin{equation}
    \mathcal{F}_{\mathrm{AM}}[\psi,\psi^*]=\int dx \left[-\frac{1}{2}\psi^*\partial_x^2\psi+\frac{1}{2}|\psi|^4-\frac{2}{3}|\psi|^3-\mu |\psi|^2\right].
\end{equation}
Therefore, the AM equation of state (energy per particle) $e_{\mathrm{AM}}(\rho)$ is obtained by setting $\psi=\sqrt{\rho}$, and we have
\begin{equation}
    e_{\mathrm{AM}}=\frac{1}{2}\rho-\frac{2}{3}\rho^{1/2}.
\end{equation}
The chemical potential is $\mu=\rho-\rho^{1/2}$ and the physical branch of the density as a function of the chemical potential is simply
\begin{equation}
    \rho(\mu)=\left[\frac{1}{2}\left(1+\sqrt{1+4\mu}\right)\right]^2.
\end{equation}
The equilibrium chemical potential is $\mu_0=-2/9$, and the chemical potential at the spinodal point is $\mu_*=-1/4$. Putting all this together, we see that the locally minimized AM grand potential $\Omega_{\mathrm{AM}}$ as a function of length ($L$) and chemical potential $\mu$, is given by
\begin{equation}
    L^{-1}\Omega_{\mathrm{AM}}(\mu)=\frac{1}{2}[\rho(\mu)]^2-\frac{2}{3}[\rho(\mu)]^{3/2}-\mu\rho(\mu).
\end{equation}
For $\mu=\mu_0$, we have $\Omega_{\mathrm{AM}}=0$, at which point the vacuuum also minimizes $\Omega_{\mathrm{AM}}$. For $\mu_*<\mu<\mu_0$, the absolute minimum occurs for $\psi\equiv 0$, that is, the vacuum. Therefore, if we are willing to describe mean-field liquids at densities that are lower than the equilibrium density, we must accept the local minimum at finite density as valid, for otherwise, the system should be in a state that is empty of particles. 

\begin{figure}[b]
    \centering
    \includegraphics[width=\columnwidth]{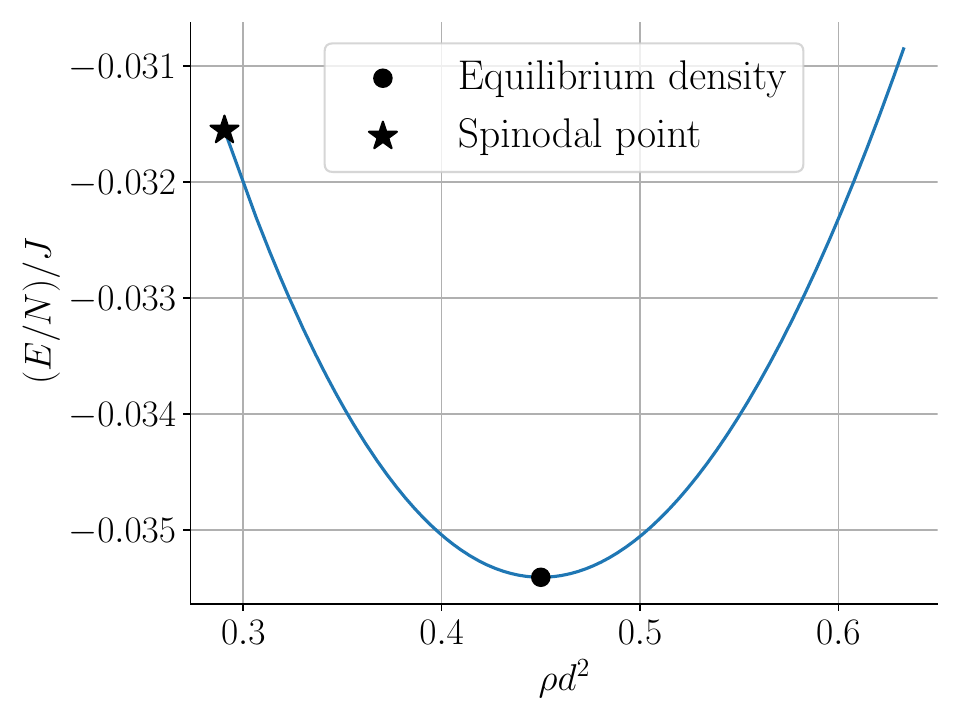}
    \caption{Equation of state of a quantum liquid obtained by means of the Gutzwiller ansatz. The parameters of Hamiltonian (\ref{eq:single_site_ham}) are $U/J = -0.33$, $W/J = 1.39$.}
    \label{fig:EoS}
\end{figure}

Proceeding, in our case, by accepting the local minimum with non-zero density as the solution, we show in Fig.~\ref{fig:EoS} the equation of state (EoS) for two particular values $U$ and $W$ of the two- and three-body interaction strengths. There, we see the formation of a liquid state, with an equilibrium density $\rho_0$ at the minimum of the energy per particle, and a spinodal point $\rho_*$, for which $\partial\mu/\partial \rho|_{\rho=\rho_*}=0$. The spinodal point marks the density below which the system is unstable against the formation of droplets. Note that, as opposed to the Gross-Pitaevskii approach (see, e.g., Ref.~\cite{Astrakharchik2018})), with the Gutzwiller Ansatz there is no solution beyond the spinodal point, i.e., no solution has density $\rho<\rho_*$. From the equation of state, we calculate the speed of sound $c$ using the familiar relation
\begin{equation}
    mc^2 = \frac{\partial \mu}{\partial N},\label{eq:mccuadrado}
\end{equation}
where the mass is to be identified with the tight-binding expression $J=1/2m$. In the next section, we calculate the speed of sound directly from the low-energy excitation spectrum and compare the results.

\section{Inhomogeneous system and low-energy excitations}
In general, the Gutzwiller ansatz allows us to explore the dynamics of non-uniform systems. With the parametrization of the ground state Ansatz in  Eq.~(\ref{eq:Gutz_gs}), we now find the set of parameters $\{a_{n}(j)\}$ that minimizes the grand potential $\Omega=H-\mu N$, with the Hamiltonian $H$ of Eq.~(\ref{eq:hamiltonian}). To do so, we have two equivalent options: (i) we apply the decoupling approximation of Eq.~(\ref{eq:decoupling}) to the grand potential; or (ii) we use directly the Gutzwiller Ansatz, Eq.~(\ref{eq:Gutz_gs}), and minimize the Lagrangian $\mathcal{L}=H-\mu N -i\partial_t$ (we set $\hbar\equiv 1$) to allow for time-dependent dynamics. 

We proceed with approach (ii), with time-dependent amplitudes $a_n(j;t)$, which gives the following set of time-dependent Gutzwiller equations ($\hbar\equiv 1$) \cite{Krutitsky2011, Zakrzewski2005}
\begin{equation}
    i\frac{da_{n}(j;t)}{dt}=\sum_{m=0}^{\infty}\Omega_j(n,m)a_{m}(j;t),
\end{equation}
where the site-dependent matrix elements $\Omega_j(n,m)$ have the form
\begin{align}
  \Omega_j(n,m) &= \Bigg[  \frac{U}{2}(n-1)   +\frac{W}{6}(n-1)(n-2) \nonumber\\
  & +(2J-\mu) \Bigg] n  \delta_{n,m} + J(\psi_{j+1}^* + \psi_{j-1}^*)\psi_j \delta_{n,m}\nonumber
 \\ &-J\sqrt{m}(\psi_{j+1}^* + \psi_{j-1}^*)\delta_{m,n+1} \nonumber \\ 
 &-J\sqrt{n}(\psi_{j+1} + \psi_{j-1})\delta_{n,m+1}.\label{eq:Gutz_eq}
\end{align}

\subsection{Low-energy excitations}
To extract low-energy excitations, we consider plane-wave perturbations moving on top of the ground state, as
\begin{equation}
a_n(j;t) = a_n^{(0)} + a_n^{(1)}(j;t) + \ldots,
\end{equation}
where $a_n^{(0)}$ is the $n$-particle amplitude of the ground state wave function, Eq.~(\ref{eq:phi}), in the homogeneous, thermodynamic limit, and $a_n^{(1)}(j;t)$ is a site- and time-dependent perturbation of the form 
\begin{equation}
a^{(1)}_{n}(j;t) = u_{k,n} e^{i(k \cdot j - \omega_k t)} + v^*_{k,n} e^{-i(k \cdot j - \omega_k t)}.
\end{equation}
Following Refs.~\cite{Krutitsky2011,Trefzger2008}, we linearize the resulting set of equations and find the following eigenvalue problem:
\begin{equation}
\label{eq:excitations}
    \omega_k\begin{pmatrix} \mathbf{u}_k \\ \mathbf{v}_k \end{pmatrix} = \begin{pmatrix}
        A_k & B_k \\ -B_k & -A_k \end{pmatrix} \begin{pmatrix} \mathbf{u}_k \\ \mathbf{v}_k \end{pmatrix},
\end{equation}
\begin{figure}[b]
    \centering
    \includegraphics[width=\columnwidth]{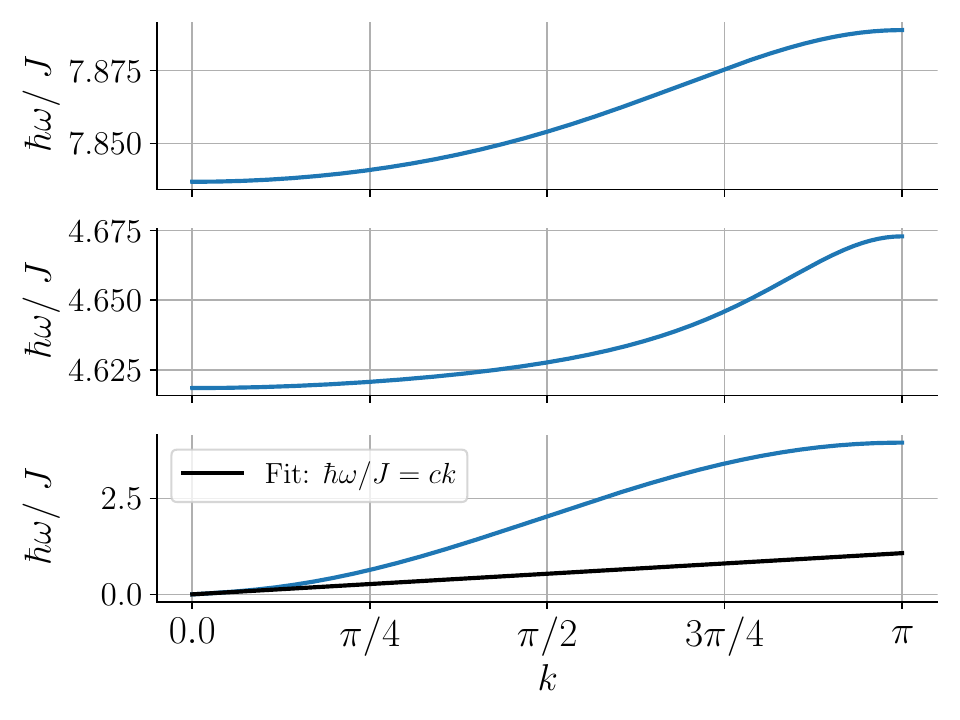}
    \caption{Low-energy excitations of the liquid. The panels represent up-to-down the third, second and first excitation branches of the system. Moreover, in the lowest panel, we show how, as we should expect, the lowest energy excitations are phononic (i.e. gapless), and therefore the speed of sound is well-defined. The parameters of Hamiltonian (\ref{eq:single_site_ham}) are $U/J = -0.33$, $W/J = 1.39$ and we used $\mu = \mu_{0}$ as defined in Fig.~\ref{fig:EoS}. The parameters of the fit are $c/J = 0.32$ and $\alpha/J^2 = 0.75$.}
    \label{fig:excitations}
\end{figure}
where $(\mathbf{u}_k,\mathbf{v}_k)$ are the complex amplitudes of the right and left moving waves in the Fock basis, while $A_k$ and $B_k$ are matrices, whose elements take the form
\begin{align}
&A_k^{nm}=-J_0\psi^{(0)}\left(\sqrt{m}\delta_{m,n+1}+\sqrt{n}\delta_{n,m+1}\right)\nonumber \\
&+\Bigg[
\frac{U}{2}n(n-1)+\frac{W}{6}n(n-1)(n-2) \nonumber \\
& +(2J-\mu) n - \hbar \omega_0
\Bigg] \delta_{m,n}\nonumber\\
&-J_k\left[
\sqrt{n+1} \sqrt{m+1} a_{n+1}^{(0)} a_{m+1}^{(0)}
+ 
\sqrt{n} \sqrt{m} a_{n-1}^{(0)} a_{m-1}^{(0)}
\right],
\end{align}
and
\begin{align}
B^{nm}_{k} = -J_k 
&\Big[ \sqrt{n+1} \sqrt{m} a_{n+1}^{(0)} a_{m-1}^{(0)}\nonumber \\
&+ \sqrt{n} \sqrt{m+1} a_{n-1}^{(0)} a_{m+1}^{(0)} \Big].
\end{align}
Similarly to $a_n^{(0)}$, the quantity $\psi^{(0)}$ is the lowest order term of the expansion of Eq.~(\ref{eq:order_parameter}) over the homogeneous ground state, that is, it is the order parameter in the thermodynamic limit. Above, we have defined $J_k = 2J(1-2\sin(k/2)^2)$, and 

\begin{align}
    &\hbar \omega_0 = 4J\psi^{(0)2}+ \sum_{n=0}\Bigg( \frac{U}{2}n(n-1) \nonumber\\
    &+\frac{W}{6}n(n-1)(n-2)+(2J-\mu) n\Bigg)\left|a_n^{(0)}\right|^2.
\end{align}

\begin{figure}[b]
    \centering
    \includegraphics[width=\columnwidth]{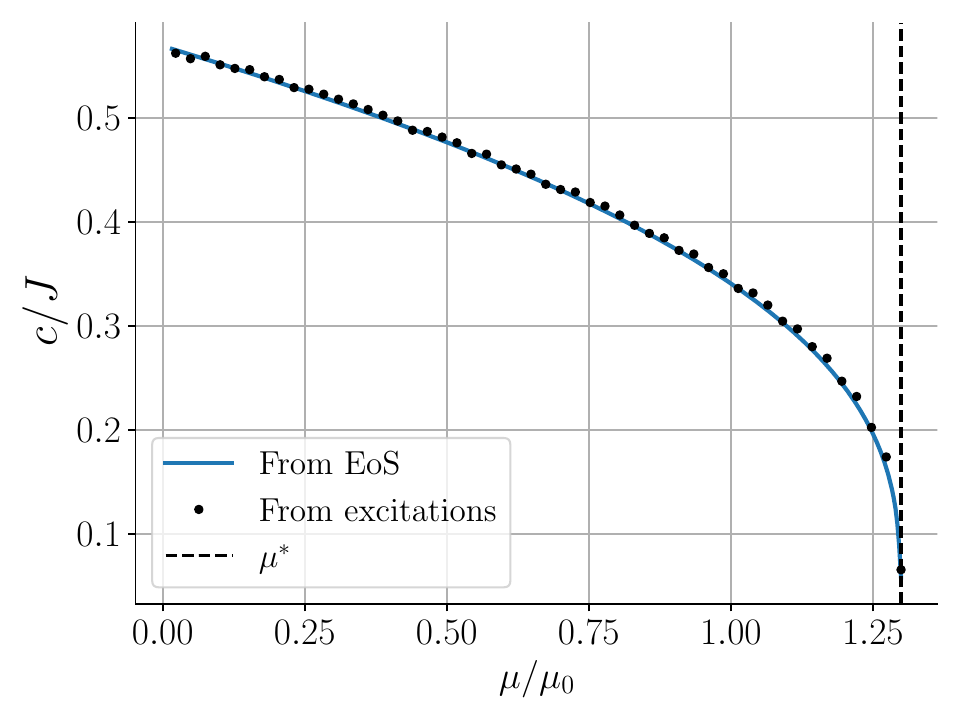}
    \caption{Speed of sound computed using Eq.~(\ref{eq:mccuadrado}) with the data from Fig.~\ref{fig:EoS} (blue solid line) and by linearly fitting the lowest branch of the excitation spectrum  (see Fig.~\ref{fig:excitations}) for small values of the momentum (dots).  The parameters of the Hamiltonian (\ref{eq:single_site_ham}) are $U/J = -0.33$, $W/J = 1.39$.}
   
    \label{fig:speedofsound}
\end{figure}

The eigenvalues and eigenvectors of the block matrix in Eq.~(\ref{eq:excitations}) give the excitation spectrum of the system in the Gutzwiller approximation, see Fig.~\ref{fig:excitations}. In particular, from the lowest (non-zero), gapless excitation branch $\omega_k^{(1)}$, we can obtain the speed of sound $c$, since, for $k\to 0$, $\omega_k^{(1)}\to c|k|$ \cite{Huber2008}, and compare it with that obtained from the equation of state, Eq.~(\ref{eq:mccuadrado}). In Fig.~\ref{fig:speedofsound}, we plot the speed of sound calculated both using the equation of state and by using a linear fit to the gapless excitation branch for low quasi-momenta, at particular values $U/J=-0.33$ and $W/J=1.39$ of the two- and three-body interaction strengths. There, we observe good agreement between the two approaches.

\subsection{Droplet states}
As we have seen, the Gutzwiller approximation supports liquid states at zero temperature. Therefore, besides the homogeneous, thermodynamic ground state, we should expect quantum droplet states with finite particle numbers to exist in the system  \cite{Luo2021,Petrov2015,Petrov2016}.

A priori, it is possible to calculate the droplet states by direct local minimization using a droplet-like wave function as a seed. We have followed this procedure and, unfortunately, convergence is very poor as the iterative, numerical procedure tends to either converge towards the vacuum of particles or the homogeneous ground state. We have therefore taken a different route, by minimizing, using an appropriate Ansatz, a density functional $E_{\mathrm{GW}}$ of the following form
\begin{equation}
    E_{\mathrm{GW}}[\rho]=\sum_{j}\left[\rho^{1/2}(j)(\hat{T}\rho^{1/2})(j)+\mathcal{E}_{\mathrm{GW}}(\rho(j))\right].\label{eq:functional}
\end{equation}
Above, $\rho(j)$ is the site-dependent density of the droplet, $\mathcal{E}_{\mathrm{GW}}(\rho)$ is the energy density obtained from the Gutzwiller approximation in the thermodynamic limit, and
\begin{equation}
    (\hat{T}f)(j)=-J(f(j+1)+f(j-1)-2f(j)),\label{eq:kinetic}
\end{equation}
is the kinetic energy term on a lattice. Equation~(\ref{eq:kinetic}) is the discrete analogue to the usual lowest-order correction term in density functional theories of liquids \cite{Dalfovo1995} \footnote{For the droplets, since their average size is much larger than the lattice spacing we have indistinctly used the usual continuous derivative. Since $J = 1/2m$ the kinetic term reads $T[\rho] 
= J\left|\frac{\text{d}\sqrt{\rho}}{\text{d}x}\right|^2$}. We use the following Ansatz \cite{Sprung1997,Morera2020b} for the droplet densities,

\begin{equation}
\label{eq:ansatz}
    \rho(j\;;\gamma,D )=\frac{N}{D} \frac{ \sinh\left(\frac{\gamma}{2}\right)}{\cosh\left( \frac{\gamma}{2}  \right) 
  + \cosh{\left(\frac{\gamma}{D} j\right)}}
\end{equation}
 
\begin{figure}[t]
    \centering
    \includegraphics[width=\columnwidth]{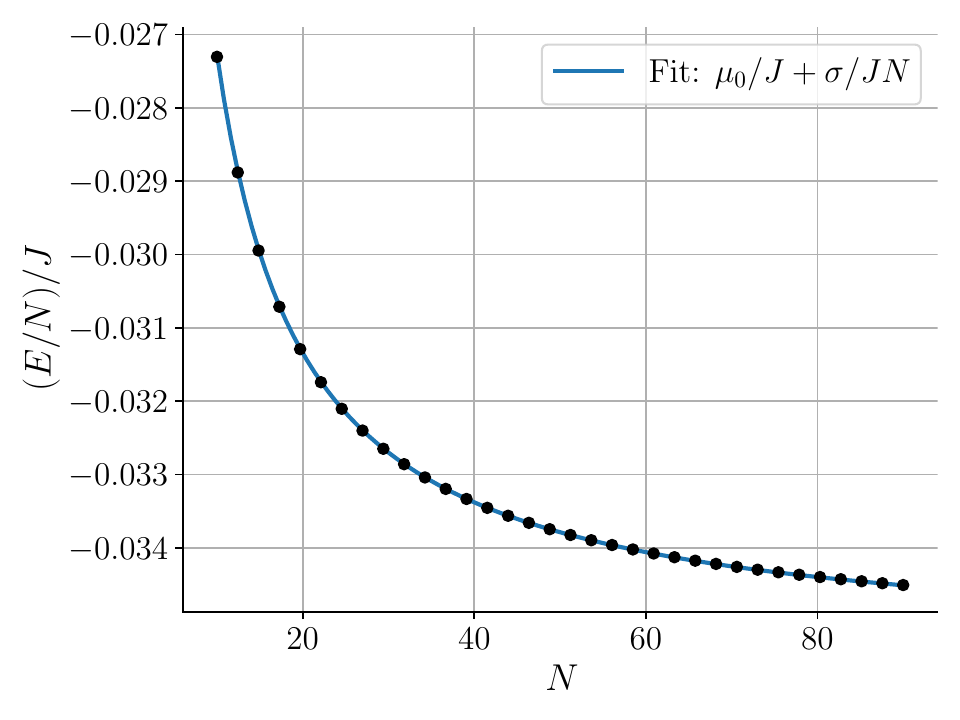}
    \caption{Energy per particle in terms of the number of particles computed using Ansatz (\ref{eq:ansatz}). We used the data from Fig.~\ref{fig:EoS} to construct $\mathcal{E}_{GW}$ as shown in Eq.~(\ref{eq:functional}). The surface energy obtained from the fit is $\sigma/J = 0.081$. The parameters of the Hamiltonian (\ref{eq:single_site_ham}) are $U/J = -0.33$, $W/J = 1.39$.}
    \label{fig:energy_droplet}
\end{figure}
\begin{figure}[b]
    \centering
    \includegraphics[width=\columnwidth]{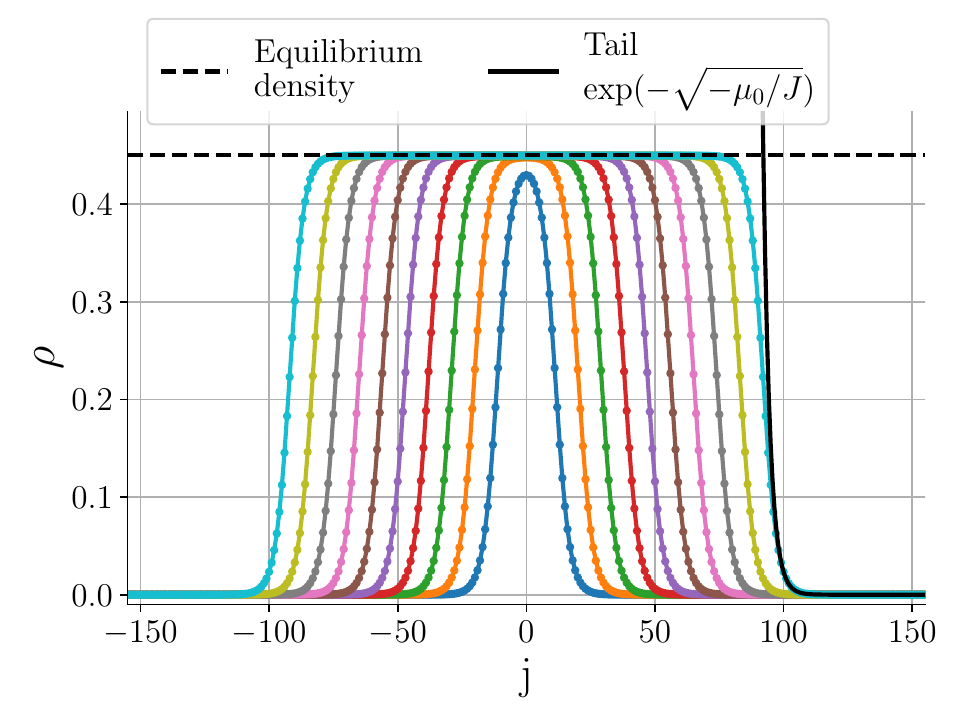}
    \caption{Droplets profiles computed using Ansatz (\ref{eq:ansatz}). We used the data from Fig.~\ref{fig:EoS} to construct $\mathcal{E}_{GW}$ as shown in Eq.~(\ref{eq:functional}). The parameters of the Hamiltonian (\ref{eq:single_site_ham}) are $U/J = -0.33$, $W/J = 1.39$. From narrowest to broadest, the droplets have, respectively, $N_i = N_0 + i\Delta N$ ($i=0,1,\ldots, 9$) particles, with $N_0 = 10$ and $\Delta N = 8.0.$}
    \label{fig:droplets}
\end{figure}

The density above has a central density given by $\rho(0) = \frac{N}{D}\frac{\sinh\left(\gamma/2\right)}{1+\cosh\left(\gamma/2\right)}$ and exponential tails given by $\rho(j) \sim \exp{\left(-\gamma  |j|/2 D\right)}$ for large $|j|$. We will later find, see Fig.~\ref{fig:droplets}, that these agree with the thermodynamic values $\rho(0) = \rho_0$ and $\gamma/D = \sqrt{-\mu/J}$  \cite{Morera2022a}.

The free parameters in Ansatz (\ref{eq:ansatz}) are $D$ and $\gamma$. Minimizing the functional (\ref{eq:functional}) with respect to the free parameters, at fixed particle number $N$, we obtain the energy per particle, central density and chemical potential as functions of the particle number in vacuum. The energy per particle, in one dimension, has the following large-$N$ asymptotics~\cite{Chin1992,Pandharipande1986,Stringari1987}
\begin{equation}
    \frac{E}{N}=\mu_0+\frac{\sigma}{N}+\frac{g(N)}{N},\label{eq:ENdroplet}
\end{equation}
where $\sigma$ is the surface energy of the droplets, and $g(N)$ is a function with $\lim_{N\to\infty}g(N)=0$ but is non-analytic in $N^{-1}$, see Fig.~\ref{fig:energy_droplet}. Therefore, the chemical potential behaves, asymptotically, as $\mu=\mu_0+g(N)$. From the density functional in Eq.~(\ref{eq:functional}), and the Ansatz in Eq.~(\ref{eq:ansatz}), it can be seen that $g(N)$ may be exponential in the number of particles (see Appendix~\ref{sec:appendix}).

\section{Conclusions}
We have investigated both ground-state properties and low-energy excitations of a Bose-Hubbard model in which, besides the conventional on-site two-body interactions we have included a contact three-body potential. This extended framework allowed us to study the formation of a self-bound quantum liquid phase building upon previous studies where only gas phases were considered~\cite{Krutitsky2011}. We made use of the Gutzwiller approximation both in its time-dependent and independent form. We first found that this approach is suitable for studying quantum liquids on a lattice, obtaining an equation of state for a system in which the competition between attractive and repulsive forces gives rise to a self-bound state. 

We have also discussed the limitations of this approach, as we have found that, as the equilibrium density is approached, the global minimum of the mean-field functional becomes the vacuum state; however, we have argued that this is common to other well-known mean-field theories such as modified Gross-Pitaevskii equations. Nonetheless, by appropriately choosing the local minimum of the functional we can reproduce the full equation of state and excitation spectrum, capturing the defining characteristics of a quantum liquid. These include phononic (gapless) low-energy excitations, a spinodal point as well as the presence of an equilibrium density where the system attains its lowest energy per particle.

\begin{figure}[t]
    \centering
    \includegraphics[width=\columnwidth]{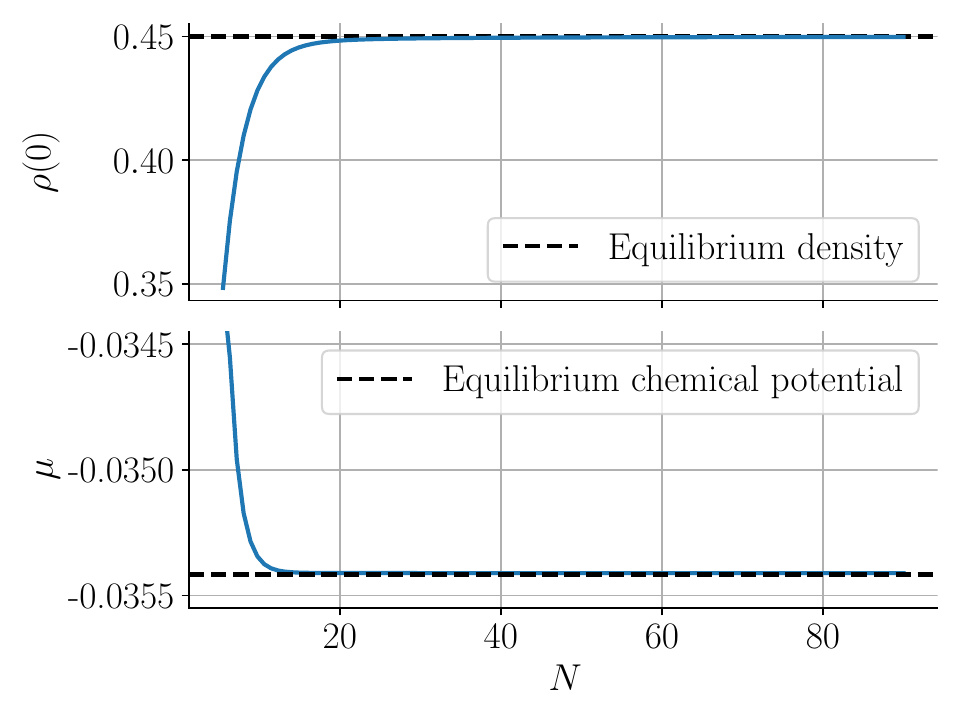}
    \caption{Central density and chemical potential in terms of the number of particles computed using Ansatz (\ref{eq:ansatz}). We used the data from Fig.~\ref{fig:EoS} to construct $\mathcal{E}_{GW}$ as shown in Eq.~(\ref{eq:functional}). The surface energy obtained from the fit is $\sigma/J = 0.081$. The parameters of the Hamiltonian (\ref{eq:single_site_ham}) are $U/J = -0.33$, $W/J = 1.39$.}
    \label{fig:properties}
\end{figure}

As a benchmark for comparing the time-dependent and time-independent approaches, we chose the speed of sound.  We computed it in the thermodynamic limit by means of Eq.~(\ref{eq:mccuadrado}) and compared it with the one obtained from the dispersion relation of gapless excitations in the long wavelength regime. The agreement found showcases the validity of our analysis.

Moreover, we have studied the formation of droplets in inhomogeneous finite systems. These self-bound states, within the context of ultracold atoms, have been considered primarily incorporating mean-field effects and the inclusion of the Lee-Huang-Yang term \cite{Skov2021,Guo2021,Lee1957} that takes into account quantum fluctuations. Here we have taken another route and studied droplets that are balanced by the competition of two-body and three-body interactions. The profiles are obtained using a variational approach. Following this procedure, we studied the energetic properties of the droplets and compared them with the predictions of the liquid drop model finding agreement between them. We have discussed the relevance of the non-analytic (possibly exponential) corrections that are present in the liquid drop model from a density functional. Since we have limited our study to a variational analysis we have not been able to access the precise functional form of these corrections. A more detailed study of the problem using DMRG or DMC method of this system, as done in Ref.~\cite{Morera2022a}, may attain a much better resolution of the properties of the droplets and provide us with another method to study their dynamical properties.

\begin{acknowledgments}
This work is supported by the Spanish Ministry of Science, Innovation and Universities through the national research and development grant PID2021-126039NA-I00, grant FPU No.~FPU22/03376 funded by MICIU/AEI/10.13039/501100011033 (D.P.-C.), and the Ram{\'o}n y Cajal Program (Grant No.~RYC2020-029961-I) (M.V.).
\end{acknowledgments}

\appendix
\section{One-dimensional liquid drop model}\label{sec:appendix}
Here, we use the lattice version of the liquid drop model in one spatial dimension and show that, within its framework, the energy per particle in a one-dimensional droplet follows Eq.~(\ref{eq:ENdroplet}), and that the function $g(N)$ may feature exponential terms. 

We begin by assuming that $N$ is large, so that $|\mu/\mu_0-1|\ll 1$, and $|\rho(0)/\rho_0-1|\ll 1$. According to the model, the large majority of particles in the droplet occupy a central plateau with density $\rho(0)\equiv \rho_c$ and diameter $D$. We then set
\begin{equation}
    N=\sum_{j=-D/2}^{D/2}\rho_c\implies D+1 = \frac{N}{\rho_c}.\label{eq:radius}
\end{equation}
 We assume that the droplet is leptodermous, so that we may model the surface energy density, $\Sigma(j)$, as
\begin{equation}
    \Sigma(j) \equiv \frac{\sigma}{2}\left[\delta_{j,D/2+1}+\delta_{j,-D/2-1}\right].
\end{equation}
If we ignore at first kinetic energy contributions, the energy of such a droplet takes the value
\begin{equation}
    E=(D+1)\mathcal{E}(\rho_c)+\sigma,
\end{equation}
where $\mathcal{E}(\rho)$ is the energy density in the thermodynamic limit. Using Eq.~(\ref{eq:radius}) to relate particle number and diameter, we obtain
\begin{equation}
    \frac{E}{N}=e(\rho_c)+\frac{\sigma}{N},
\end{equation}
with $e(\rho)$ the energy per particle (equation of state) in the thermodynamic limit. The chemical potential, on the other hand, is independent of the surface energy in one spatial dimension, since
\begin{equation}
    \mu(N)=\frac{\partial\mathcal{E}(\rho_c)}{\partial{\rho_c}},
\end{equation}
where $\rho_c=\rho_c(N)$.
Calling $\delta\rho=\rho_c-\rho_0$, from the liquid drop model without kinetic terms, we would simply have, following \cite{Treiner1986}, $\partial_N(\delta\rho)=0$, which is special for one spatial dimension. However, this result, within this framework, only means that $\delta\rho(N)$ is not an analytic function of a power of $N^{-1}$. Non-analytic deviations, such as exponentials, are not included. Kinetic energy contributions, negligible in higher dimensions, are exponential. Therefore, we must consider them in one dimension because of the absence of more dominant, polynomial contributions. To see this, we may use Ansatz (\ref{eq:ansatz}) and minimize the grand-canonical functional. The correction of the droplet's chemical potential due to the kinetic energy is given by 
\begin{align}
    -J\frac{\partial_x^2\rho^{1/2}}{\rho^{1/2}}\Big|_{x=0}&\approx J\left(\frac{\gamma}{D}\right)^2\frac{e^{-\gamma/2}}{2}\nonumber\\
    &\approx -\frac{\mu_0}{2}\exp\left({-\frac{\sqrt{-\mu_0/J}}{2\rho_0}}N+C\right),\label{eq:correction}
\end{align}
where $C$ is a constant that accounts for the inaccuracy in setting $D=N/\rho_0$ in the liquid drop model. This indicates that $g(N)$ in Eq.~(\ref{eq:ENdroplet}) may be exponential in the number of particles, see Fig.~\ref{fig:properties}. However, the exponent is not necessarily correct, since Eq.~(\ref{eq:correction}) is a local quantity, which is notoriously difficult to reproduce accurately without access to the exact density profile. Moreover, these exponential terms add up to other exponential terms that come from the thermodynamic contribution $\mu=\mu(\rho_c)$. In the particular case of Ref.~\cite{Astrakharchik2018}, this is exactly the case, and the kinetic energy contribution cancels out.

\bibliography{bibliography}
\end{document}